\documentclass[11pt]{article}

\usepackage[T1]{fontenc}
\usepackage[utf8]{inputenc}
\usepackage[british]{babel}
\usepackage[margin=1in]{geometry}
\usepackage{graphicx}
\usepackage{float}
\usepackage{mathtools}
\usepackage{amsmath}
\usepackage{amsfonts}
\usepackage{amssymb}
\usepackage{booktabs}
\usepackage{multirow}
\usepackage{array}
\usepackage[table]{xcolor}
\usepackage{tikz}
\usepackage{needspace}
\usepackage[authoryear,round]{natbib}
\usepackage[parfill]{parskip}
\usepackage[hidelinks]{hyperref}

\graphicspath{{figures/}}
\title{An Evidence-Aware Framework for EEG Microstate Analysis:\\
Improved Sensitivity to Alzheimer's Disease and Ageing}

\author{%
\begin{minipage}{0.94\textwidth}\centering
Kaidong Wu\textsuperscript{1},
Haili Ye\textsuperscript{1},
Ptolemaios G Sarrigiannis\textsuperscript{2,3},
Daniel J Blackburn\textsuperscript{3}, and
Fei He\textsuperscript{1,*}\\[0.75em]
\small \textsuperscript{1}Centre for Computational Science and Mathematical Modelling, Coventry University, Coventry, United Kingdom\\
\small \textsuperscript{2}Academic Department of Neurosciences, Sheffield Teaching Hospitals NHS Foundation Trust, Sheffield, United Kingdom\\
\small \textsuperscript{3}Sheffield Institute for Translational Neuroscience (SITraN), University of Sheffield, Sheffield, United Kingdom\\[0.4em]
\small \textsuperscript{*}Corresponding author: Fei He, Centre for Computational Science and Mathematical Modelling, Coventry University, Coventry CV1 2TL, United Kingdom; \href{mailto:fei.he@coventry.ac.uk}{fei.he@coventry.ac.uk}
\end{minipage}}\date{}

\begin{document}
\maketitle

\begin{abstract}
Electroencephalography (EEG) microstate analysis commonly converts each scalp topography into a winner-take-all hard label and summarises the resulting sequence using duration, occurrence, coverage, transition structure, and symbolic complexity. Although useful and interpretable, this hard label readout discards template evidence strength, assignment ambiguity, and periods of low confidence. Here, we introduce a template evidence trajectory framework for EEG topographic state dynamics. We sampled scalp topographies at local maxima of Global Field Power (GFP), the across-electrode standard deviation of voltage and a measure of instantaneous field strength. For each sampled topography, we computed its evidence for all templates or subject-specific topographic communities and treated the conventional hard label as a compressed readout of this multivariate trajectory. We then derived two evidence-aware extensions of classical hard label descriptors. High-evidence episode duration and episode rate quantify the temporal clustering or fragmentation of each state’s strongest evidence periods. Null state Lempel-Ziv complexity (LZC) explicitly encodes time points with insufficient template evidence.

We evaluated this framework across four resting-state EEG datasets spanning Alzheimer’s disease and age-related variation across adulthood. The fixed state models were \(K\)-means, atomize and agglomerate hierarchical clustering (AAHC), and hidden Markov models (HMMs), each evaluated at \(K=4\) and \(K=7\). We also included adaptive, subject-specific graph communities identified by Leiden and Infomap. At \(K=4\), group-level templates aligned with canonical A--D patterns, whereas \(K=7\) provided a finer-grained repertoire. Both settings showed high global explained variance across datasets. Across 32 dataset-model cases, trajectory-derived duration effects exceeded their matched hard label counterparts in all cases at the representative setting and in 30--32 cases across the wider parameter grid. Null state LZC improved over traditional LZC in most cases, although the gain was smaller and more dependent on state model granularity. Classification analyses showed modest but consistent gains for trajectory or combined feature sets.

These results suggest that retaining template evidence provides a more sensitive readout of EEG topographic state dynamics while preserving compatibility with conventional microstate analysis.
\end{abstract}


\section{Introduction}

Electroencephalographic (EEG) microstate analysis provides a compact description of rapid, large-scale changes in the scalp voltage pattern during spontaneous brain activity. It represents the continuous EEG as a sequence of brief periods during which this pattern remains relatively stable. The recurring patterns are called microstates. Each microstate is a spatial voltage configuration across electrodes and is thought to reflect the momentary engagement of a large-scale functional network. A small set of representative topographies, called templates, is estimated from the data, and their temporal expression is summarised using interpretable descriptors \cite{khanna2015microstates,michel2018eeg}. Classical temporal parameters include duration, occurrence, and coverage or occupancy, often complemented by transition structure and symbolic sequence complexity. These metrics have been widely applied in cognitive and clinical EEG research, and large-scale reliability studies suggest that several classical microstate measures exhibit good reliability \cite{khanna2014reliability,kleinert2024reliability}. Most conventional analyses use a winner-take-all readout: each observed EEG topography is assigned to the most similar template, producing a discrete hard label sequence from which temporal descriptors are computed.

Although the hard label sequence is useful and easy to interpret, it loses information about the template evidence underlying each assignment. Interpreting hard labels as physiological state visits assumes that the winning template is both adequately supported and dominant. This means that a time point with one clearly dominant template and another with nearly tied template evidence may receive the same label. The first assignment is made with high confidence, whereas the second is weak or ambiguous. Once the evidence profile is collapsed to its argmax label, the strength of the winner, its margin over competing templates, and whether the evidence is mixed or weak are no longer recoverable from the label sequence. This limitation aligns with methodological work questioning a strictly discrete interpretation of EEG microstates. Geometric analyses have described microstates as spatially and temporally continuous phenomena \cite{mishra2020eeg}. Probabilistic approaches have explicitly modelled assignment uncertainty \cite{dinov2017modeling}. Recent reviews have also emphasised that continuous representations may complement symbolic microstate sequences \cite{haydock2025eeg}.

These critiques have largely focused on state representation and assignment uncertainty. Their implications for the downstream statistics used in most empirical microstate studies remain insufficiently examined. Mean duration, occurrence, coverage, and sequence complexity are still typically calculated after the evidence profile has been reduced to a hard segmentation. Even when graded topography-to-template similarity or activation time courses are available, they have most often been used as auxiliary fitting measures, visualisation outputs, or EEG-informed regressors rather than as the basis for redefining classical temporal biomarkers \cite{britz2010bold,xu2020eeg,nagabhushan2024microstatelab}. The open question is whether the evidence discarded by winner-take-all assignment can support matched, evidence-aware extensions of conventional duration- and complexity-based descriptors. A related question is whether these extensions provide reproducible gains in sensitivity across different datasets and state model constructions.

Among descriptors derived from hard label microstate sequences, two families are central to the present work. The first comprises duration, occurrence, and coverage, which describe how long a state persists, how often it appears, and how much of the recording it occupies \cite{croce2020eeg,da2020eeg}. Coverage is naturally compatible with the winner-take-all framework, because it explicitly measures the fraction of recording time for which a template is the best-fitting state. Duration and occurrence, however, depend on how the hard sequence partitions time into state visits. A contiguous hard label segment may reflect sustained and dominant template evidence, or merely a sequence of weak winner-take-all assignments in which competing templates remain similarly plausible. Conventional duration therefore measures persistence of the winning label, but it cannot distinguish persistence supported by strong evidence from ambiguous persistence. Occurrence has a related ambiguity: for a given state,
\[
\mathrm{occurrence}_k
=
\frac{\mathrm{coverage}_k}
{\mathrm{mean\ duration}_k},
\]
increased occurrence can reflect greater state prevalence, shorter and more fragmented visits, or both. Thus, classical temporal parameters are jointly informative, but they do not isolate the temporal organisation of strongly supported template evidence.

The second family concerns symbolic sequence complexity. Lempel--Ziv complexity (LZC) is commonly applied to microstate sequences because it quantifies the diversity of recurring symbolic subsequences without assuming a first-order Markov process \cite{tait2020eeg,von2024complexity,wan2024beyond}. However, LZC inherits the assumption that every hard symbol represents a meaningful state visit. Under winner-take-all assignment, topographies that are weakly explained by all available templates are nevertheless forced into the nearest state and thus become indistinguishable from assignments made with high confidence. The resulting symbolic complexity may therefore reflect not only the temporal organisation of well-supported states, but also arbitrary labels introduced during periods of low template evidence.

To address this gap, we use the full topography-to-template evidence time course as the primary readout of a topographic state model. We refer to this multivariate time course as a template evidence trajectory. Let \(e_k(t)\geq 0\) denote the evidence of the EEG topography at time \(t\) for state \(k\). We parameterise the normalised trajectory using an evidence-sharpening exponent \(g\):
\[
p_k^{(g)}(t)
=
\frac{e_k(t)^g}
{\sum_{j=1}^{K}e_j(t)^g},
\qquad g\geq 1.
\]
At \(g=1\), the trajectory preserves the directly normalised relative evidence across states. Increasing \(g\) emphasises the dominant states while progressively suppressing weaker competitors; in the limiting case, the representation approaches a one-hot winner-take-all readout. The parameter \(g\) therefore defines a controlled continuum between graded and increasingly discrete state representations. This allows us to test whether evidence-aware effects depend on the degree of evidence sharpening rather than on a single normalisation choice. The conventional hard label remains
\[
z(t)=\arg\max_k p_k^{(g)}(t),
\]
whereas the full vector \(p^{(g)}(t)\) retains the relative support for all states.

Using these trajectories, we identify high-evidence episodes and derive confidence-conditioned descriptors of state persistence and fragmentation. For each state and each value of \(g\), high-evidence samples are identified using within-state quantile thresholds, and consecutive above-threshold samples form an episode. Because the threshold is defined from each state’s own evidence distribution, the marginal fraction of samples considered high-evidence is standardised across states. Episode duration and episode rate therefore summarise how a fixed upper-tail fraction of template evidence is organised in time: sustained episodes indicate temporally clustered high-confidence evidence, whereas many brief episodes indicate fragmentation. These measures are not intended to replace conventional coverage, which remains an interpretable measure of winner-state prevalence. Instead, they complement coverage by characterising the temporal organisation of strongly supported state evidence after controlling the amount of high-evidence samples considered.

We apply the same evidence-aware principle to symbolic complexity by introducing null state LZC. The conventional LZC algorithm is retained, but the symbolic sequence is modified so that time points failing an evidence-confidence criterion are assigned a dedicated null symbol rather than being forced into the nearest template. Null state LZC therefore distinguishes periods of low template evidence from genuine state visits while preserving the established symbolic-complexity framework. Together, high-evidence episode metrics and null state LZC extend two common classes of hard label descriptors, namely state persistence and sequence complexity, without changing the underlying state templates.

We evaluated the framework on four resting-state EEG datasets to test two types of variation: Alzheimer’s disease (AD) and age-related variation across adulthood. These settings were chosen because both dementia and ageing are known to modulate resting-state EEG organisation and have repeatedly been associated with altered microstate dynamics. In AD and related cognitive impairment, previous studies have reported changes in microstate duration, occurrence, coverage, and transition patterns, including shortened microstate duration as well as associations with disease progression, amyloid burden, and diagnostic status \cite{strik1997decreased,dierks1997eeg,musaeus2019microstates,lian2021altered,yan2024abnormal,zhang2025abnormalities}. Ageing provides a complementary validation setting because it reflects graded inter-individual differences rather than a binary diagnostic contrast. Previous EEG and magnetoencephalography (MEG) studies have reported age-related differences in microstate templates and temporal features across development and adulthood \cite{jabes2021resting,bagdasarov2022spatiotemporal,hill2023eeg,huang2024magnetoencephalography}. We therefore used AD versus Healthy Control (HC), i.e. AD--HC, comparisons and age-association analyses as two tests of whether evidence-aware descriptors enhance sensitivity to meaningful variation in EEG topographic state dynamics.

We also asked whether the findings generalised across state model constructions. For this purpose, we evaluated both conventional fixed state models and adaptive subject-specific models. Conventional models included fixed-\(K\) clustering or latent state approaches such as \(K\)-means, atomize and agglomerate hierarchical clustering (AAHC), and hidden Markov models (HMMs). In parallel, recent electrophysiological state modelling work has increasingly used individualised and graph-based representations of brain dynamics \cite{varley2022network}. Related EEG and MEG studies have applied recurrence or dynamic-connectivity graphs with community detection to identify electrophysiological meta-states \cite{nunez2021abnormal,nunez2025altered,sandonis2025disrupted}. Motivated by this adaptive perspective, we represented scalp topographies sampled at local maxima of global field power (GFP) as nodes in subject-specific similarity graphs. GFP is the across-electrode standard deviation of voltage and summarises the strength of the instantaneous scalp field. We then used Leiden or Infomap to identify individualised topographic communities \cite{traag2019louvain,rosvall2008maps}. These communities were interpreted as subject-specific topographic states rather than as members of a shared canonical class set.

For every state model, we compared matched hard label and template evidence readouts derived from the same underlying templates or communities. Our primary analyses tested whether high-evidence episode duration and episode rate yielded stronger disease- and age-related effects than conventional duration and occurrence, and whether null state LZC improved upon ordinary hard label LZC. We further evaluated whether these effects generalised across evidence-sharpening values \(g\), quantile thresholds, datasets, and state model constructions. We also tested whether they translated into improved classification or regression performance. We hypothesised that conventional hard labels would remain useful summaries of state prevalence and persistence of the winning label, but that retaining graded template evidence would provide a more sensitive measure of EEG topographic state dynamics.

\begin{figure}[H]
    \centering
    \includegraphics[width=\textwidth]{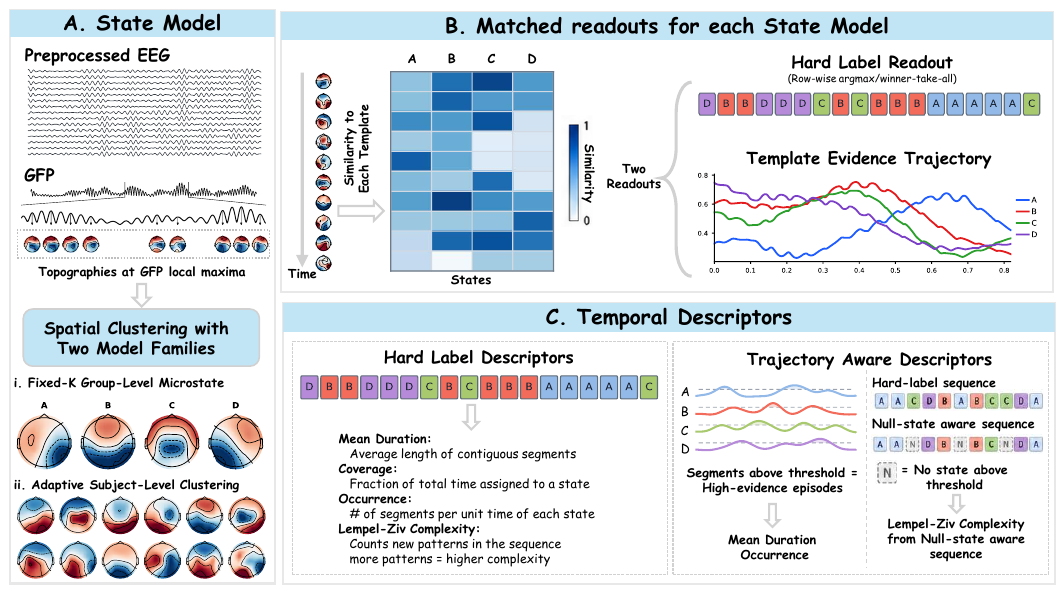}
    \caption{
Overview of the proposed analysis framework. 
(A) Preprocessed EEG was first summarised at local maxima of global field power (GFP), which is the across-electrode standard deviation of voltage. The corresponding scalp topographies were used to construct state models. Two state model families were considered: (i) fixed-$K$ group-level microstate models, which produce aligned group-level templates across participants, and (ii) adaptive subject-level clustering models, which allow the number and form of states to vary across individuals. 
(B) For each state model, every time point (or GFP peak) was represented by its similarity to all templates, yielding a template-similarity matrix. From this matched representation, two readouts were derived: a conventional hard label sequence obtained by winner-take-all assignment (row-wise argmax), and a continuous template evidence trajectory that preserves graded similarity to all templates over time. 
(C) Temporal descriptors were then extracted from each readout. From the hard label sequence, conventional descriptors such as mean duration, coverage, occurrence, and Lempel--Ziv complexity (LZC) were computed. From the trajectory representation, trajectory-aware descriptors were constructed in two ways: thresholded high-evidence segments were used to define trajectory-based duration and occurrence, and a symbolic sequence with an explicit null state was used to compute LZC when no template exceeded the evidence threshold. This framework enables direct comparison between matched hard label and evidence-aware temporal descriptors across different state model constructions.
    }
    \label{fig:pipeline}
\end{figure}

\section{Material and Methods}

\subsection{Datasets and EEG preprocessing}

We evaluated the proposed EEG topographic state readouts on four eyes-closed resting-state EEG datasets spanning two validation settings: AD--HC group differences and age-related variation across adulthood. First, we used OpenNeuro ds004504, a routine clinical EEG dataset containing eyes-closed recordings from patients with AD, patients with frontotemporal dementia (FTD), and HC \citep{Miltiadous2023DS004504}. For the primary AD--HC analysis, we included the 36 AD participants and 29 HC participants and excluded the FTD group. Second, we used the Chung-Ang University Hospital EEG (CAUEEG) cohort \citep{kim2023deep}. To obtain a cleaner diagnostic contrast, AD was restricted to participants annotated as early-onset or late-onset AD, whereas HC was restricted to normal-control labels; subjective memory impairment, vascular dementia, mixed dementia, frontotemporal dementia, and other non-AD clinical labels were excluded. This yielded 228 AD and 272 HC participants. Third, we used OpenNeuro ds005385, a lifespan resting-state EEG dataset from the Dortmund Vital Study \citep{Getzmann2024DS005385}. We used the session-1 pre-task eyes-closed resting-state recording, yielding 608 healthy adults aged 20--70 years. Fourth, we used the Two Decades-Brainclinics Research Archive for Insights in Neurophysiology (TD-BRAIN), a large clinical EEG archive with resting-state recordings and diagnostic metadata \citep{vanDijk2022TDBRAIN}. For the age-association analysis, we included adult participants ($\geq 18$ years) from the eyes-closed resting-state condition whose indication and formal diagnostic status were both healthy, yielding 179 participants with available recordings.

All analyses used eyes-closed resting-state EEG. Non-EEG channels, including trigger/status, electrooculography (EOG), electrocardiography (ECG), and electromyography (EMG) channels, were excluded before topographic analysis. For ds004504, CAUEEG, and ds005385, EEG was analysed at 200 Hz. TD-BRAIN was analysed at 500 Hz after applying the dataset's official preprocessing workflow to recordings stored in BioSemi Data Format (BDF/BDF+). Across datasets, alpha-band activity was isolated using a zero-phase Butterworth band-pass filter from 8 to 13 Hz, followed by linear detrending when required and common-average referencing. For TD-BRAIN, we retained a fixed-duration eyes-closed segment for each participant after official preprocessing, rather than defining subject-specific variable-length excerpts based on artefact annotations. No cross-dataset channel interpolation or spatial harmonisation was applied; all topographic analyses were performed in the native scalp montage of each dataset.

Global field power (GFP) was computed as the across-channel standard deviation of the alpha-band, common-average-referenced signal. This scalar summarises the instantaneous strength of the scalp electric field. Local maxima of GFP were detected using a minimum inter-peak distance of 10 ms. For each GFP peak, the corresponding scalp topography was extracted and used as the primary sample for state model construction. Each peak was assigned a temporal weight using midpoint interpolation between adjacent GFP peaks, so that the peak represented the time interval extending halfway to the preceding and following peaks. These peak-wise temporal weights were used for all duration-based descriptors.

\subsection{Topographic state model construction}

All topographic state models were constructed from scalp topographies at GFP local maxima. For subject \(s\), let
\[
X_s=\{x_{s,1},x_{s,2},\dots,x_{s,N_s}\}
\]
denote the GFP peak topographies, where \(x_{s,i}\in\mathbb{R}^{C}\) is the scalp map over \(C\) EEG channels at the \(i\)-th GFP peak. GFP peaks were used because they provide relatively stable, high signal-to-noise topographic samples and correspond to the standard sampling strategy in EEG microstate analysis.

We considered two complementary families of state models. The first family followed the conventional fixed-\(K\), group-template microstate framework. This family represents the classical microstate-analysis setting in which a shared set of template maps is estimated at the group level and then backfitted to each participant. At \(K=4\), these templates can be related to canonical classes, whereas \(K=7\) provides a finer-grained repertoire. Specifically, we fitted \(K\)-means, atomize and agglomerate hierarchical clustering (AAHC), and hidden Markov models (HMMs) at both \(K=4\) and \(K=7\) using a two-level group clustering procedure. Each subject was first fitted independently at the specified \(K\), producing subject-level candidate maps. These subject-level maps were then pooled across participants and clustered again to estimate a common group-level template set. The resulting group templates were finally backfitted to each subject's GFP peak sequence. This procedure follows the standard logic of group-level microstate analysis: it provides shared state identities across participants while reducing the influence of subjects with more GFP peaks on the final group templates.

The second family comprised adaptive subject-level state models. These models relaxed the assumption that all participants are optimally described by the same number of states and the same group-level template repertoire. For each subject, GFP peak topographies were represented as nodes in a subject-specific topographic similarity graph,
\[
G_s=(V_s,E_s,A_s),
\]
where each node corresponded to a GFP peak map and edge weights encoded polarity-invariant topographic similarity. Community detection was then applied separately within each subject. We evaluated Leiden, a modularity-based community detection method, and Infomap, an information-theoretic flow-based method. These adaptive models produced subject-specific state repertoires with subject-specific numbers of states \(K_s\), and therefore were not interpreted as canonical microstates.

Using both model families, we tested whether the proposed evidence trajectory framework generalised beyond a single state modelling assumption. The fixed-\(K\) models evaluated the framework in the classical microstate setting with shared group templates and interpretable state-specific descriptors. The adaptive models evaluated the same evidence-aware readouts under subject-specific state repertoires, where state identities need not be aligned across participants. Thus, fixed-\(K\) and adaptive models served complementary roles: the former anchored the analysis in conventional microstate methodology, whereas the latter tested robustness to recent concerns about imposing a common group-level template structure on heterogeneous participants.

\subsection{Matched hard label and evidence trajectory readouts}

For each fitted state model, we computed two matched readouts from the same underlying templates or communities: a conventional hard label sequence and a continuous evidence trajectory. For fixed-\(K\) group models, the templates were the two-level group templates backfitted to each subject. For adaptive Leiden and Infomap models, each community was represented by a subject-specific prototype map.

Template evidence was computed as polarity-invariant topographic similarity between a GFP peak map and each template. In the primary implementation, both maps and templates were shape-normalised before computing absolute spatial correlation. The absolute value treated maps that differed only in sign as equivalent, so the evidence score reflected topographic alignment rather than absolute amplitude:
\[
E_{s,k}(i)=\left|\mathrm{corr}\left(x_{s,i},\mu_{s,k}\right)\right|.
\]
The evidence vector was then sharpened and normalised across states:
\[
P^{(g)}_{s,k}(i)=
\frac{(E_{s,k}(i)+\epsilon)^g}
{\sum_{j=1}^{K_s}(E_{s,j}(i)+\epsilon)^g},
\]
where \(g\) is an evidence-sharpening parameter. \(g=1\) preserves the original relative evidence, whereas larger values make the strongest state more dominant and reduce the contribution of weaker candidate states.

For fixed-\(K\) group models, the hard label readout was obtained by winner-take-all backfitting to the group templates. For adaptive graph models, the hard label readout was the native Leiden or Infomap community assignment. The evidence trajectory readout retained the full vector \(P^{(g)}_{s}(i)\) across all states. Thus, every evidence-aware descriptor was compared with a hard label descriptor derived from the same state model.

\subsection{Hard label and evidence-aware descriptors}

These matched readouts were then used to compute conventional and evidence-aware temporal descriptors. The hard label sequence supplied the classical measures, whereas the full template evidence vector supplied their evidence-aware counterparts.

For a given subject \(s\), let \(z_s(i)\) denote the hard label state assigned to GFP peak \(i\), and let \(P^{(g)}_{s,k}(i)\) denote the sharpened and normalised evidence for state \(k\). The parameter \(g\) controls evidence sharpening: \(g=1\) preserves the original relative evidence, whereas larger values make the strongest state more dominant. All descriptors were computed on the GFP peak sequence using midpoint temporal weights \(w_{s,i}\), so that each peak represented the time interval bounded by neighbouring GFP peaks. 

Conventional hard label temporal descriptors were computed following standard microstate definitions. Mean duration was defined as the average length of contiguous same-state segments, occurrence as the number of same-state segments per unit time, and coverage as the fraction of total recording time assigned to a given state. State-specific duration, occurrence, and coverage were computed only for fixed-\(K\) group templates, where state identities were aligned across subjects. For adaptive Leiden and Infomap models, we reported label-invariant global descriptors because their state repertoires were subject-specific.

Evidence-aware duration and occurrence were computed from high-evidence episodes. For state \(k\), evidence-sharpening parameter \(g\), and percentile threshold \(p\), we defined a subject- and state-specific evidence threshold
\[
\theta_{s,k}^{(g,p)}
=
Q_p\left(P^{(g)}_{s,k}(i)\right),
\]
where \(Q_p\) is the weighted within-subject percentile of that state's evidence trajectory. A high-evidence episode was a contiguous interval satisfying
\[
P^{(g)}_{s,k}(i) > \theta_{s,k}^{(g,p)}.
\]
Episodes shorter than the minimum-duration criterion were removed using the same duration constraint as in the hard label analysis. Trajectory mean duration was computed as the mean duration of these high-evidence episodes, pooled across states for global descriptors or computed separately by state for fixed-\(K\) state-specific analyses. Trajectory occurrence was defined as the corresponding high-evidence episode rate per unit time.

Trajectory-based duration and hard label duration capture different properties. Hard label duration measures how long the winning label persists, regardless of whether the winning state is strongly or weakly supported. Evidence-trajectory duration instead measures how long high-confidence evidence for a state persists. Because the threshold is defined within each subject and state, the descriptor focuses on the temporal organisation of strong evidence rather than on absolute state prevalence.

Symbolic sequence complexity was quantified using normalised Lempel--Ziv complexity (LZC). Conventional LZC was computed directly from the hard label sequence,
\[
LZC^{hard}_{s}
=
LZC\left(z_s(1),z_s(2),\dots,z_s(T_s)\right).
\]
This measures the diversity of symbolic subsequences in the hard label stream, but it assumes that every time point is meaningfully assigned to one of the available states.

To make sequence complexity confidence-aware, we computed a null state LZC. First, each time point was assigned to the state with the highest within-state percentile rank of evidence, so that each state's evidence was evaluated relative to its own subject-specific distribution. Let \(\tilde{z}_s(i)\) denote this percentile-selected label. The selected label was retained only if its sharpened evidence exceeded the corresponding state-specific threshold; otherwise, the time point was assigned to a null state \(N\):
\[
z^{null}_{s}(i)
=
\begin{cases}
\tilde{z}_s(i), &
P^{(g)}_{s,\tilde{z}_s(i)}(i) >
\theta_{s,\tilde{z}_s(i)}^{(g,p)},\\
N, & \text{otherwise}.
\end{cases}
\]
Null LZC was then computed using the same normalised LZC algorithm applied to this null state-aware sequence,
\[
LZC^{null}_{s}(g,p)
=
LZC\left(z^{null}_{s}(1),z^{null}_{s}(2),\dots,z^{null}_{s}(T_s)\right).
\]
Thus, null LZC preserved the conventional symbolic-complexity framework while explicitly encoding periods of low confidence that are invisible in an ordinary winner-take-all label sequence.

In summary, the hard label descriptors quantify the temporal organisation of discrete winner-take-all states, whereas the evidence-aware descriptors quantify persistence and complexity after retaining confidence information from the full evidence trajectory. All comparisons were matched within the same state model, allowing differences between hard label and evidence-aware descriptors to be attributed to the readout strategy rather than to the underlying state model construction.

\subsection{Statistical analysis and validation}
All analyses were conducted at the subject level and organised around two validation axes: AD--HC group differences and age-related variation across adulthood. For statistical effect-size analyses, AD--HC datasets were evaluated using Welch tests and Cohen's $d$, whereas ageing datasets were evaluated using Pearson's correlation with chronological age.

Predictive validation was performed using subject-native features to avoid information leakage. In repeated 5-fold cross-validation, fitting group-level microstate templates separately within every training fold would require repeatedly re-estimating the full state model pipeline, whereas fitting templates once on the full cohort would allow held-out test subjects to influence feature construction. Therefore, classification analyses used subject-level state models: each subject's GFP peak topographies were clustered and summarised independently, and matched hard label and evidence-aware descriptors were computed from the same subject-specific model.

For AD-related datasets, models classified AD cases versus HC. For the two lifespan datasets, age-related classification was defined a priori by contrasting younger adults below 35 years with older adults above 60 years. Classification performance was estimated using repeated 5-fold cross-validation and summarised by balanced accuracy and the area under the receiver operating characteristic curve (ROC-AUC). Scaling, model fitting, and hyperparameter tuning were performed within the training folds only, with held-out folds used exclusively for evaluation.

Each evidence-aware descriptor was compared against a matched hard label baseline from the same state model family. Thus, subject-level KMeans, AAHC, HMM, Leiden, and Infomap evidence-derived features were evaluated against their corresponding hard label features, and combined models concatenated the two feature families. Threshold and parameter variants were treated as predefined sensitivity analyses rather than post hoc searches on held-out data.

\section{Results}\label{sec:Results}

\subsection{Group-level templates recover canonical structure at K=4 and a finer repertoire at K=7}

\begin{figure}[H]
    \centering
    \includegraphics[width=\textwidth]{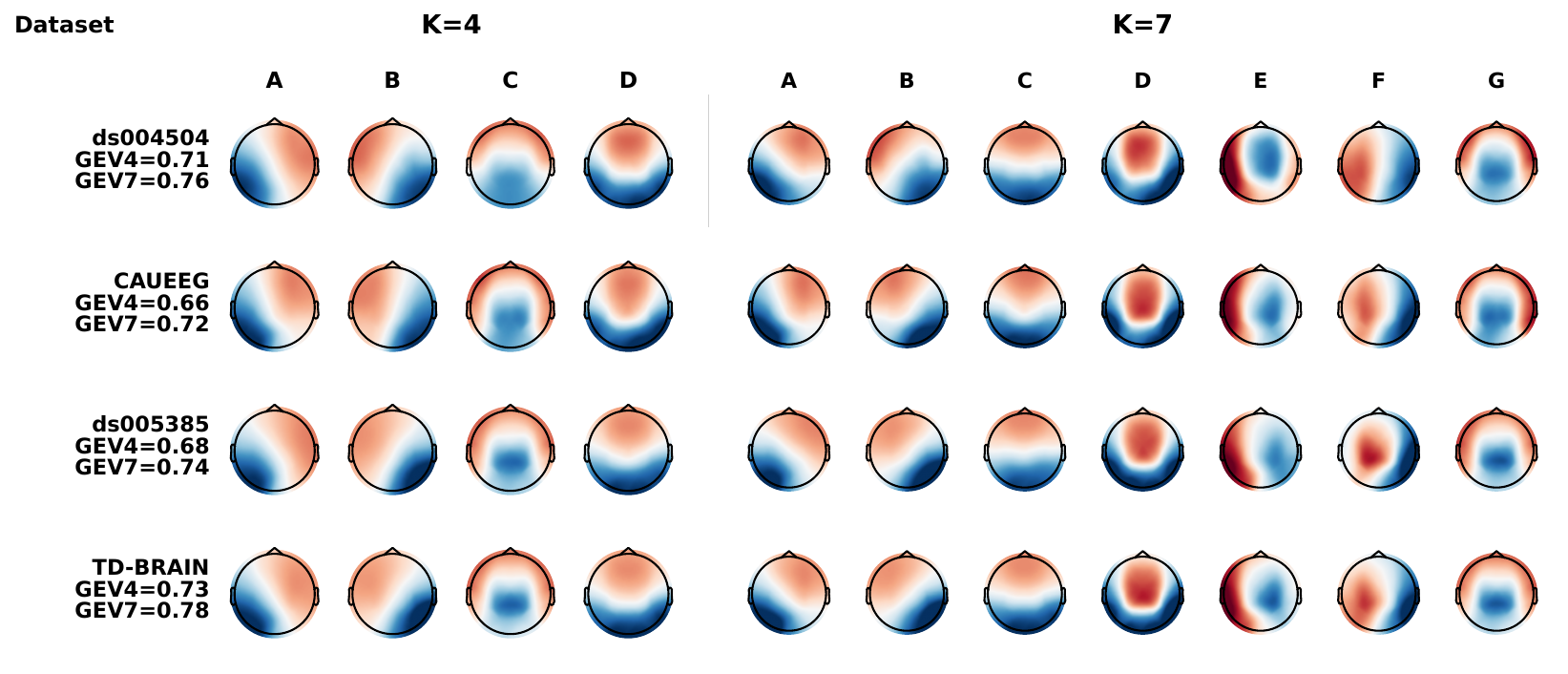}
   \caption{
Group-level reference templates obtained with the fixed-$K$ microstate branch across the four datasets. For each dataset, templates are shown for $K=4$ and $K=7$, together with global explained variance (GEV), the proportion of topographic variance accounted for by the templates. At $K=4$, the templates show clear correspondence with canonical microstate-like patterns A--D across datasets. Increasing to $K=7$ yields a more fine-grained set of topographies while preserving the overall canonical spatial structure and increasing GEV.
}
    \label{fig:group_templates}
\end{figure}

We first checked the conventional state model by inspecting group-level templates across the four datasets, using \(K\)-means as a representative fixed-\(K\) method. When \(K=4\), the fitted templates showed clear correspondence with the widely validated canonical A--D microstate patterns, including a lateralised A/B pair and more anterior--posterior or symmetric C/D-like templates (Figure~\ref{fig:group_templates}). Global explained variance (GEV), the proportion of topographic variance accounted for by the fitted templates, was consistently high across datasets, ranging from 0.66 to 0.73. When \(K=7\), the templates provided a more fine-grained repertoire while preserving canonical-like spatial structure, with GEV increasing to 0.72--0.78. These results indicate that the conventional group-level branch produced stable and interpretable reference templates across datasets.

Leiden and Infomap were not evaluated using canonical template alignment because they produced subject-specific adaptive topographic communities. These individualised states were used for label-invariant summary descriptors and matched hard label versus template evidence readout comparisons, rather than for cross-subject canonical-class interpretation.

\subsection{Template evidence descriptors show stronger statistical effects than matched hard label descriptors}
We next compared the statistical effects obtained from matched hard label and template evidence descriptors. In this analysis, we focused on global descriptors, including global mean duration and LZC, rather than state-specific measures tied to individual fixed-\(K\) templates. Global descriptors summarise the overall temporal organisation of the topographic state sequence and are not driven by any particular state identity. They therefore provide a common label-invariant readout across different fixed-$K$ settings and the adaptive Leiden and Infomap models.

For each descriptor family, we summarised performance across 32 dataset--model cases, defined by four datasets and eight state model constructions: \(K\)-means, AAHC, and HMM at \(K=4\) and \(K=7\), plus Leiden and Infomap. Table~\ref{tab:gp-matrix} reports the mean effect-size ratio
\[
R=\frac{|\mathrm{effect}_{traj}|}{|\mathrm{effect}_{hard}|},
\]
together with the number of cases in which the trajectory descriptor exceeded the matched hard label descriptor.

For the duration family, template evidence descriptors showed a consistent advantage across the full parameter grid. Mean \(R\) ranged from 1.34 to 1.44 across tested \(g\) and percentile-threshold combinations. The largest mean ratio was observed at \(g=1\) and percentile 50, with \(R=1.44\), and the trajectory descriptor exceeded the hard label descriptor in all 32 cases. Similar ratios were observed across neighbouring parameter settings. The duration-family improvement was therefore not restricted to a single threshold or sharpening value.

Null state LZC also showed an overall advantage over ordinary hard label LZC, although the gain was smaller and more model-dependent. Across tested parameters, mean \(R\) ranged from 1.13 to 1.23. The maximum mean ratio of \(R=1.23\) was reached at several parameter settings; at the highlighted setting of \(g=7\) and percentile 75, null state LZC exceeded hard label LZC in 29 of 32 cases. Thus, both global descriptor families favoured the evidence-aware readout, with the strongest and most consistent effect observed for duration-related descriptors.

Dataset- and model-specific effects are shown in Table~\ref{tab:gp-matrix}. At the representative duration setting (\(g=1\), percentile 50), the trajectory duration descriptor produced larger absolute effects than hard label duration in every dataset and state model construction. For example, in ds004504, \(K\)-means at \(K=4\) increased from \(d=-0.871\) for hard label duration to \(d=-1.778\) for trajectory duration. In CAUEEG, the same comparison increased from \(d=-0.486\) to \(d=-0.655\). For age-association datasets, trajectory duration also yielded stronger correlations, increasing from \(r=-0.172\) to \(r=-0.252\) in ds005385 and from \(r=-0.228\) to \(r=-0.345\) in TD-BRAIN for \(K\)-means at \(K=4\).

For LZC, null state LZC generally increased the effect magnitude relative to ordinary hard label LZC. The improvement was most consistent for fixed state models. For example, in ds004504, AAHC at \(K=4\) increased from \(d=1.147\) for hard label LZC to \(d=1.824\) for null state LZC. In CAUEEG, AAHC at \(K=4\) increased from \(d=0.748\) to \(d=0.919\). In the age datasets, null state LZC also increased the effect for most fixed state models. The gain was smaller for adaptive Leiden and Infomap models, where ordinary hard label LZC was already strong in some datasets.

We did not include state-specific duration and occurrence analyses in this global descriptor summary. Such measures are most interpretable for group-level fixed-\(K\) templates, where state labels are aligned across participants; in contrast, adaptive Leiden and Infomap models yield subject-specific state identities and are better compared using pooled or label-invariant global descriptors.

\subsection{Trajectory and combined features improve classification performance}

We then evaluated whether the descriptor differences translated into subject-level classification performance. Table~\ref{tab:subject-level-classification} summarises repeated 5-fold outer cross-validation results. The two AD datasets were evaluated as AD versus HC classification tasks. The two age datasets were evaluated using a young/old contrast (\(\leq 35\) vs. \(\geq 60\) years). Hard label and trajectory feature sets each contained 16 descriptors, and the combined model used all 32 descriptors.

Across datasets, the best balanced accuracy was obtained by either the trajectory feature set or the combined feature set. In ds004504, balanced accuracy increased from \(0.806\pm0.096\) for hard label features to \(0.828\pm0.081\) for trajectory features and \(0.839\pm0.091\) for the combined model. The ROC--AUC was highest for the combined model (\(0.888\pm0.083\)), compared with \(0.862\pm0.089\) for hard label features. In CAUEEG, balanced accuracy increased from \(0.704\pm0.041\) for hard label features to \(0.729\pm0.039\) for trajectory features and \(0.739\pm0.039\) for the combined model. ROC--AUC increased from \(0.778\pm0.042\) for hard label features to \(0.790\pm0.040\) for trajectory features and \(0.814\pm0.037\) for the combined model.

For the age-contrast datasets, trajectory features also improved performance relative to hard label features. In ds005385, balanced accuracy increased from \(0.689\pm0.058\) for hard label features to \(0.696\pm0.057\) for trajectory features and \(0.703\pm0.055\) for the combined model. ROC--AUC was highest for trajectory features (\(0.777\pm0.060\)). In TD-BRAIN, trajectory features produced the largest improvement, with balanced accuracy increasing from \(0.701\pm0.095\) to \(0.752\pm0.098\), and ROC--AUC increasing from \(0.800\pm0.097\) to \(0.831\pm0.095\). Overall, trajectory-derived descriptors showed modest but consistent predictive gains, and combined hard label plus trajectory features often provided the best or near-best performance.

\providecommand{\bestcell}[1]{%
  \begingroup
  \setlength{\fboxsep}{1.5pt}%
  \fcolorbox{black}{yellow!25}{#1}%
  \endgroup
}
\begin{table*}[t]
\centering
\caption{Effect-size ratio across trajectory parameters for global mean duration and null LZC. Each cell reports $R=|\mathrm{effect}_{\mathrm{traj}}|/|\mathrm{effect}_{\mathrm{hard}}|$, followed by the number of cases in which the trajectory descriptor exceeded the hard label descriptor out of 32 total cases. The 32 cases comprise four datasets and eight state model constructions: KMeans, AAHC, and HMM at $K=4$ and $K=7$, plus Leiden and Infomap. The largest mean $R$ within each descriptor is highlighted.}
\label{tab:gp-matrix}
\scriptsize
\setlength{\tabcolsep}{2.7pt}
\renewcommand{\arraystretch}{1.16}
\begin{minipage}[t]{0.492\textwidth}
\centering
\textbf{Mean duration}\\[0.7ex]
\resizebox{\linewidth}{!}{%
\begin{tabular}{l*{7}{c}}
\toprule
$g \backslash p$ & 50 & 55 & 60 & 65 & 70 & 75 & 80 \\
\midrule
1 & \bestcell{\shortstack{\textbf{1.44}\\\textbf{(32/32)}}} & \shortstack{1.44\\(32/32)} & \shortstack{1.43\\(31/32)} & \shortstack{1.42\\(31/32)} & \shortstack{1.41\\(31/32)} & \shortstack{1.38\\(31/32)} & \shortstack{1.34\\(32/32)} \\
2 & \shortstack{1.41\\(30/32)} & \shortstack{1.42\\(30/32)} & \shortstack{1.41\\(30/32)} & \shortstack{1.40\\(30/32)} & \shortstack{1.40\\(30/32)} & \shortstack{1.38\\(31/32)} & \shortstack{1.36\\(31/32)} \\
3 & \shortstack{1.40\\(30/32)} & \shortstack{1.40\\(30/32)} & \shortstack{1.40\\(30/32)} & \shortstack{1.39\\(30/32)} & \shortstack{1.38\\(30/32)} & \shortstack{1.38\\(31/32)} & \shortstack{1.36\\(31/32)} \\
4 & \shortstack{1.40\\(30/32)} & \shortstack{1.40\\(30/32)} & \shortstack{1.39\\(30/32)} & \shortstack{1.39\\(30/32)} & \shortstack{1.38\\(30/32)} & \shortstack{1.38\\(31/32)} & \shortstack{1.36\\(31/32)} \\
5 & \shortstack{1.39\\(30/32)} & \shortstack{1.39\\(30/32)} & \shortstack{1.39\\(30/32)} & \shortstack{1.38\\(30/32)} & \shortstack{1.38\\(30/32)} & \shortstack{1.37\\(31/32)} & \shortstack{1.36\\(31/32)} \\
6 & \shortstack{1.39\\(30/32)} & \shortstack{1.39\\(30/32)} & \shortstack{1.38\\(30/32)} & \shortstack{1.38\\(30/32)} & \shortstack{1.38\\(30/32)} & \shortstack{1.37\\(31/32)} & \shortstack{1.36\\(31/32)} \\
7 & \shortstack{1.39\\(30/32)} & \shortstack{1.39\\(30/32)} & \shortstack{1.38\\(30/32)} & \shortstack{1.38\\(30/32)} & \shortstack{1.38\\(30/32)} & \shortstack{1.37\\(31/32)} & \shortstack{1.36\\(31/32)} \\
8 & \shortstack{1.39\\(30/32)} & \shortstack{1.39\\(30/32)} & \shortstack{1.38\\(30/32)} & \shortstack{1.38\\(30/32)} & \shortstack{1.38\\(30/32)} & \shortstack{1.37\\(31/32)} & \shortstack{1.36\\(31/32)} \\
9 & \shortstack{1.39\\(30/32)} & \shortstack{1.39\\(30/32)} & \shortstack{1.38\\(30/32)} & \shortstack{1.38\\(30/32)} & \shortstack{1.38\\(30/32)} & \shortstack{1.37\\(31/32)} & \shortstack{1.36\\(31/32)} \\
10 & \shortstack{1.39\\(30/32)} & \shortstack{1.39\\(30/32)} & \shortstack{1.38\\(30/32)} & \shortstack{1.38\\(30/32)} & \shortstack{1.38\\(30/32)} & \shortstack{1.37\\(31/32)} & \shortstack{1.36\\(31/32)} \\
\bottomrule
\end{tabular}
}
\end{minipage}\hfill
\begin{minipage}[t]{0.492\textwidth}
\centering
\textbf{Null LZC}\\[0.7ex]
\resizebox{\linewidth}{!}{%
\begin{tabular}{l*{7}{c}}
\toprule
$g \backslash p$ & 50 & 55 & 60 & 65 & 70 & 75 & 80 \\
\midrule
1 & \shortstack{1.18\\(27/32)} & \shortstack{1.20\\(27/32)} & \shortstack{1.19\\(26/32)} & \shortstack{1.20\\(29/32)} & \shortstack{1.23\\(29/32)} & \shortstack{1.20\\(28/32)} & \shortstack{1.13\\(25/32)} \\
2 & \shortstack{1.17\\(26/32)} & \shortstack{1.17\\(27/32)} & \shortstack{1.19\\(27/32)} & \shortstack{1.22\\(28/32)} & \shortstack{1.23\\(29/32)} & \shortstack{1.23\\(28/32)} & \shortstack{1.17\\(26/32)} \\
3 & \shortstack{1.16\\(27/32)} & \shortstack{1.16\\(26/32)} & \shortstack{1.19\\(26/32)} & \shortstack{1.21\\(28/32)} & \shortstack{1.22\\(28/32)} & \shortstack{1.23\\(29/32)} & \shortstack{1.18\\(28/32)} \\
4 & \shortstack{1.16\\(27/32)} & \shortstack{1.15\\(27/32)} & \shortstack{1.18\\(26/32)} & \shortstack{1.20\\(26/32)} & \shortstack{1.21\\(28/32)} & \shortstack{1.23\\(29/32)} & \shortstack{1.18\\(28/32)} \\
5 & \shortstack{1.15\\(27/32)} & \shortstack{1.15\\(27/32)} & \shortstack{1.18\\(26/32)} & \shortstack{1.20\\(26/32)} & \shortstack{1.21\\(28/32)} & \shortstack{1.23\\(29/32)} & \shortstack{1.19\\(28/32)} \\
6 & \shortstack{1.15\\(26/32)} & \shortstack{1.15\\(26/32)} & \shortstack{1.18\\(26/32)} & \shortstack{1.20\\(25/32)} & \shortstack{1.20\\(28/32)} & \shortstack{1.23\\(29/32)} & \shortstack{1.19\\(28/32)} \\
7 & \shortstack{1.14\\(26/32)} & \shortstack{1.15\\(26/32)} & \shortstack{1.17\\(26/32)} & \shortstack{1.20\\(26/32)} & \shortstack{1.20\\(28/32)} & \bestcell{\shortstack{\textbf{1.23}\\\textbf{(29/32)}}} & \shortstack{1.19\\(29/32)} \\
8 & \shortstack{1.14\\(26/32)} & \shortstack{1.15\\(26/32)} & \shortstack{1.18\\(26/32)} & \shortstack{1.20\\(26/32)} & \shortstack{1.20\\(28/32)} & \shortstack{1.23\\(29/32)} & \shortstack{1.18\\(29/32)} \\
9 & \shortstack{1.14\\(27/32)} & \shortstack{1.15\\(26/32)} & \shortstack{1.17\\(26/32)} & \shortstack{1.20\\(26/32)} & \shortstack{1.20\\(28/32)} & \shortstack{1.23\\(29/32)} & \shortstack{1.18\\(29/32)} \\
10 & \shortstack{1.15\\(26/32)} & \shortstack{1.15\\(26/32)} & \shortstack{1.17\\(26/32)} & \shortstack{1.19\\(26/32)} & \shortstack{1.20\\(28/32)} & \shortstack{1.23\\(29/32)} & \shortstack{1.18\\(29/32)} \\
\bottomrule
\end{tabular}
}
\end{minipage}
\end{table*}

\providecommand{\winnercell}[1]{\begingroup\bfseries\boldmath #1\endgroup}
\begin{table*}[t]
\centering
\caption{Global mean duration and Lempel--Ziv complexity effects. AD/HC datasets report Cohen's $d$ from Welch tests; ageing datasets report Pearson's $r$ with age. Winners within each hard label versus trajectory/null LZC comparison are bolded.}
\label{tab:global-duration-lzc-effects}
\scriptsize
\setlength{\tabcolsep}{3.0pt}
\renewcommand{\arraystretch}{1.13}
\resizebox{0.97\textwidth}{!}{%
\begin{tabular}{llcccc}
\toprule
Dataset & Algorithm & Hard duration & Traj. duration & LZC & Null LZC \\
\midrule
\multirow{8}{*}{ds004504} & KMeans K=4 & $d=-0.871$ ($7.1\times10^{-4}$) & \winnercell{$d=-1.778$ ($1.8\times10^{-9}$)} & $d=1.151$ ($1.2\times10^{-5}$) & \winnercell{$d=1.771$ ($5.4\times10^{-10}$)} \\
 & KMeans K=7 & $d=-0.772$ (0.003) & \winnercell{$d=-1.717$ ($4.0\times10^{-9}$)} & $d=1.230$ ($5.7\times10^{-6}$) & \winnercell{$d=1.615$ ($6.1\times10^{-9}$)} \\
 & AAHC K=4 & $d=-1.057$ ($5.4\times10^{-5}$) & \winnercell{$d=-1.751$ ($2.5\times10^{-9}$)} & $d=1.147$ ($1.1\times10^{-5}$) & \winnercell{$d=1.824$ ($2.8\times10^{-10}$)} \\
 & AAHC K=7 & $d=-1.100$ ($3.7\times10^{-5}$) & \winnercell{$d=-1.729$ ($3.6\times10^{-9}$)} & $d=1.249$ ($2.9\times10^{-6}$) & \winnercell{$d=1.654$ ($3.3\times10^{-9}$)} \\
 & HMM K=4 & $d=-0.858$ (0.001) & \winnercell{$d=-1.734$ ($1.8\times10^{-9}$)} & $d=1.304$ ($2.3\times10^{-6}$) & \winnercell{$d=1.418$ ($1.5\times10^{-7}$)} \\
 & HMM K=7 & $d=-0.736$ (0.004) & \winnercell{$d=-1.716$ ($5.1\times10^{-9}$)} & $d=1.188$ ($8.4\times10^{-6}$) & \winnercell{$d=1.557$ ($1.5\times10^{-8}$)} \\
 & Leiden & $d=-1.384$ ($1.1\times10^{-6}$) & \winnercell{$d=-1.696$ ($7.4\times10^{-9}$)} & \winnercell{$d=1.809$ ($3.2\times10^{-10}$)} & $d=1.719$ ($2.0\times10^{-9}$) \\
 & Infomap & $d=-1.405$ ($8.8\times10^{-7}$) & \winnercell{$d=-1.696$ ($9.5\times10^{-9}$)} & $d=1.566$ ($1.4\times10^{-8}$) & \winnercell{$d=1.785$ ($7.3\times10^{-10}$)} \\
\addlinespace
\multirow{8}{*}{CAUEEG} & KMeans K=4 & $d=-0.486$ ($4.7\times10^{-8}$) & \winnercell{$d=-0.655$ ($9.9\times10^{-13}$)} & $d=0.692$ ($5.7\times10^{-14}$) & \winnercell{$d=0.875$ ($2.4\times10^{-20}$)} \\
 & KMeans K=7 & $d=-0.464$ ($2.7\times10^{-7}$) & \winnercell{$d=-0.651$ ($1.3\times10^{-12}$)} & $d=0.605$ ($4.8\times10^{-11}$) & \winnercell{$d=0.927$ ($3.0\times10^{-22}$)} \\
 & AAHC K=4 & $d=-0.567$ ($2.3\times10^{-10}$) & \winnercell{$d=-0.694$ ($4.7\times10^{-14}$)} & $d=0.748$ ($4.1\times10^{-16}$) & \winnercell{$d=0.919$ ($3.0\times10^{-22}$)} \\
 & AAHC K=7 & $d=-0.456$ ($4.4\times10^{-7}$) & \winnercell{$d=-0.680$ ($1.4\times10^{-13}$)} & $d=0.613$ ($2.6\times10^{-11}$) & \winnercell{$d=1.004$ ($1.8\times10^{-25}$)} \\
 & HMM K=4 & $d=-0.380$ ($2.3\times10^{-5}$) & \winnercell{$d=-0.632$ ($5.2\times10^{-12}$)} & $d=0.580$ ($3.0\times10^{-10}$) & \winnercell{$d=0.921$ ($1.7\times10^{-22}$)} \\
 & HMM K=7 & $d=-0.507$ ($1.9\times10^{-8}$) & \winnercell{$d=-0.657$ ($8.1\times10^{-13}$)} & $d=0.650$ ($1.7\times10^{-12}$) & \winnercell{$d=0.910$ ($1.3\times10^{-21}$)} \\
 & Leiden & $d=-0.563$ ($1.1\times10^{-9}$) & \winnercell{$d=-0.650$ ($9.5\times10^{-13}$)} & $d=0.817$ ($4.7\times10^{-18}$) & \winnercell{$d=0.885$ ($9.3\times10^{-21}$)} \\
 & Infomap & $d=-0.379$ ($3.6\times10^{-5}$) & \winnercell{$d=-0.667$ ($2.8\times10^{-13}$)} & $d=0.769$ ($3.2\times10^{-16}$) & \winnercell{$d=0.779$ ($1.3\times10^{-16}$)} \\
\addlinespace
\multirow{8}{*}{ds005385} & KMeans K=4 & $r=-0.172$ ($2.0\times10^{-5}$) & \winnercell{$r=-0.252$ ($2.9\times10^{-10}$)} & $r=0.235$ ($4.2\times10^{-9}$) & \winnercell{$r=0.275$ ($4.7\times10^{-12}$)} \\
 & KMeans K=7 & $r=-0.183$ ($5.7\times10^{-6}$) & \winnercell{$r=-0.249$ ($4.5\times10^{-10}$)} & $r=0.261$ ($6.6\times10^{-11}$) & \winnercell{$r=0.304$ ($1.7\times10^{-14}$)} \\
 & AAHC K=4 & $r=-0.207$ ($2.5\times10^{-7}$) & \winnercell{$r=-0.263$ ($4.1\times10^{-11}$)} & $r=0.261$ ($5.8\times10^{-11}$) & \winnercell{$r=0.307$ ($9.3\times10^{-15}$)} \\
 & AAHC K=7 & $r=-0.194$ ($1.4\times10^{-6}$) & \winnercell{$r=-0.254$ ($2.2\times10^{-10}$)} & $r=0.242$ ($1.6\times10^{-9}$) & \winnercell{$r=0.275$ ($5.6\times10^{-12}$)} \\
 & HMM K=4 & $r=-0.178$ ($1.0\times10^{-5}$) & \winnercell{$r=-0.259$ ($9.0\times10^{-11}$)} & $r=0.242$ ($1.6\times10^{-9}$) & \winnercell{$r=0.287$ ($5.6\times10^{-13}$)} \\
 & HMM K=7 & $r=-0.180$ ($8.2\times10^{-6}$) & \winnercell{$r=-0.251$ ($3.5\times10^{-10}$)} & $r=0.262$ ($5.2\times10^{-11}$) & \winnercell{$r=0.289$ ($3.4\times10^{-13}$)} \\
 & Leiden & $r=-0.224$ ($2.4\times10^{-8}$) & \winnercell{$r=-0.249$ ($4.8\times10^{-10}$)} & \winnercell{$r=0.307$ ($9.8\times10^{-15}$)} & $r=0.286$ ($6.9\times10^{-13}$) \\
 & Infomap & $r=-0.209$ ($2.0\times10^{-7}$) & \winnercell{$r=-0.252$ ($2.9\times10^{-10}$)} & \winnercell{$r=0.290$ ($2.8\times10^{-13}$)} & $r=0.280$ ($2.2\times10^{-12}$) \\
\addlinespace
\multirow{8}{*}{TD-BRAIN} & KMeans K=4 & $r=-0.228$ (0.002) & \winnercell{$r=-0.345$ ($2.2\times10^{-6}$)} & $r=0.316$ ($1.7\times10^{-5}$) & \winnercell{$r=0.413$ ($9.4\times10^{-9}$)} \\
 & KMeans K=7 & $r=-0.298$ ($5.0\times10^{-5}$) & \winnercell{$r=-0.349$ ($1.7\times10^{-6}$)} & $r=0.344$ ($2.4\times10^{-6}$) & \winnercell{$r=0.424$ ($3.3\times10^{-9}$)} \\
 & AAHC K=4 & $r=-0.256$ ($5.3\times10^{-4}$) & \winnercell{$r=-0.340$ ($3.2\times10^{-6}$)} & $r=0.311$ ($2.2\times10^{-5}$) & \winnercell{$r=0.395$ ($4.5\times10^{-8}$)} \\
 & AAHC K=7 & $r=-0.276$ ($1.9\times10^{-4}$) & \winnercell{$r=-0.347$ ($1.9\times10^{-6}$)} & $r=0.333$ ($5.2\times10^{-6}$) & \winnercell{$r=0.412$ ($1.0\times10^{-8}$)} \\
 & HMM K=4 & $r=-0.259$ ($4.7\times10^{-4}$) & \winnercell{$r=-0.348$ ($1.8\times10^{-6}$)} & $r=0.317$ ($1.5\times10^{-5}$) & \winnercell{$r=0.393$ ($5.4\times10^{-8}$)} \\
 & HMM K=7 & $r=-0.270$ ($2.5\times10^{-4}$) & \winnercell{$r=-0.352$ ($1.4\times10^{-6}$)} & $r=0.336$ ($4.3\times10^{-6}$) & \winnercell{$r=0.392$ ($5.8\times10^{-8}$)} \\
 & Leiden & $r=-0.281$ ($1.4\times10^{-4}$) & \winnercell{$r=-0.298$ ($5.2\times10^{-5}$)} & $r=0.391$ ($6.4\times10^{-8}$) & \winnercell{$r=0.403$ ($2.3\times10^{-8}$)} \\
 & Infomap & $r=-0.291$ ($7.8\times10^{-5}$) & \winnercell{$r=-0.295$ ($6.2\times10^{-5}$)} & $r=0.379$ ($1.7\times10^{-7}$) & \winnercell{$r=0.429$ ($2.0\times10^{-9}$)} \\
\bottomrule
\end{tabular}}
\end{table*}

\begin{figure}[H]
    \centering
    \begin{tikzpicture}\node[anchor=south west,inner sep=0] (figthree) at (0,0) {\includegraphics[width=\textwidth]{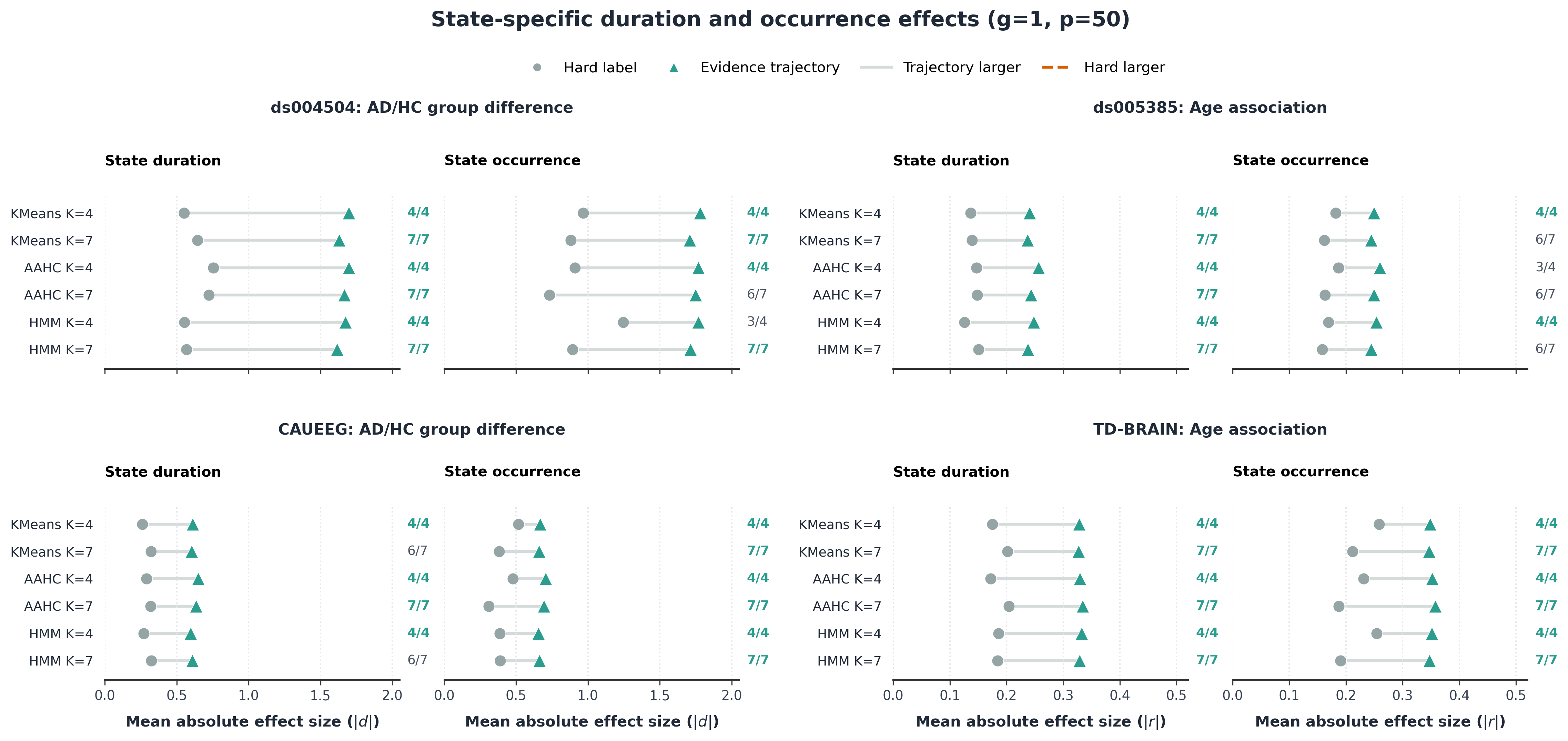}};\begin{scope}[x={(figthree.south east)},y={(figthree.north west)}]\fill[white] (0.655,0.890) rectangle (0.748,0.945);\end{scope}\end{tikzpicture}
   \caption{
Comparison of state-specific duration and occurrence effects between matched hard label and evidence trajectory descriptors at the representative setting $g=1$ and percentile threshold $p=50$. Results are shown for fixed-$K$ models across the four datasets: ds004504 and CAUEEG (AD/HC group difference, measured by mean absolute Cohen's $|d|$), and ds005385 and TD-BRAIN (age association, measured by mean absolute correlation $|r|$). Within each model, grey circles denote hard label effects and green triangles denote trajectory-based effects; connecting lines indicate paired comparisons. Fractions on the right summarise the number of states for which the trajectory descriptor produced a larger effect than the matched hard label descriptor.
}
    \label{fig:dumb_bell}
\end{figure}

\begin{figure}[H]
    \centering
    \includegraphics[width=\textwidth]{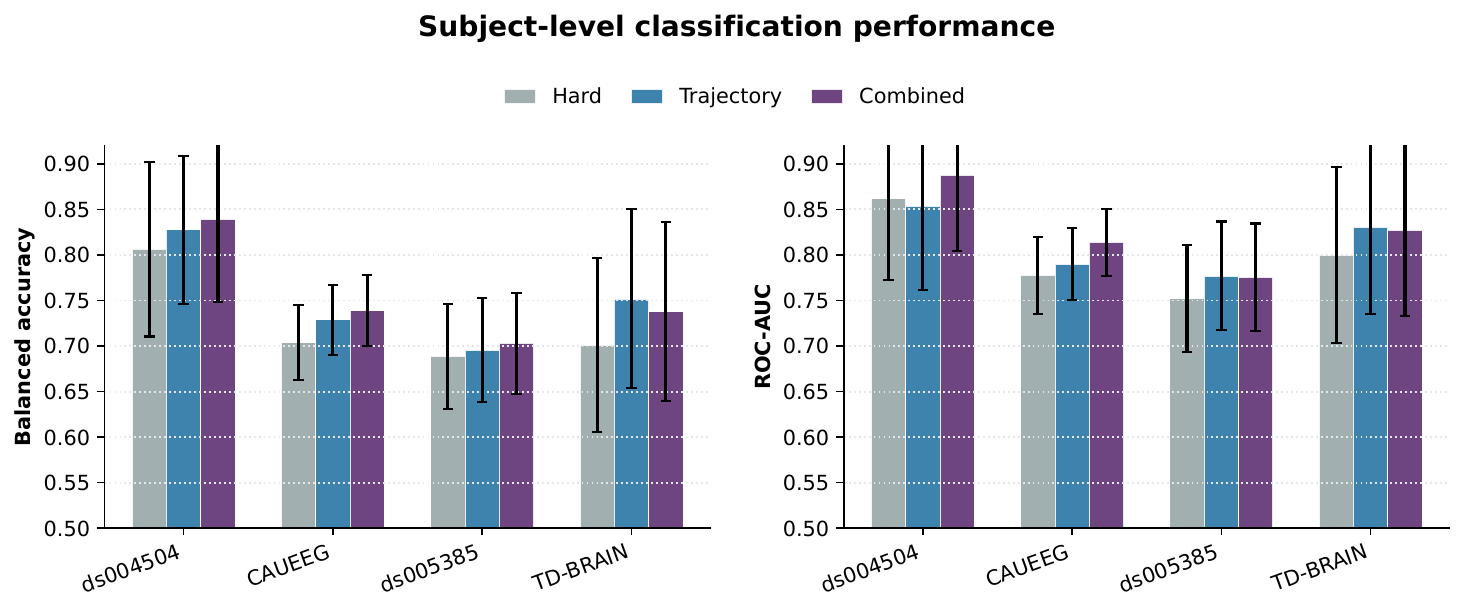}
   \caption{
Subject-level classification performance across four datasets using global duration and Lempel--Ziv complexity features. 
Bars show mean balanced accuracy and ROC-AUC across repeated cross-validation folds, and error bars indicate one standard deviation. 
Hard models used conventional hard label global duration and LZC features, trajectory models used evidence trajectory global duration and null LZC features, and combined models concatenated both feature families. 
AD/HC classification was evaluated for ds004504 and CAUEEG, while ds005385 and TD-BRAIN were evaluated as young-versus-old classification tasks using age thresholds of $\leq 35$ and $\geq 60$ years. 
}
    \label{fig:classification}
\end{figure}

\begin{table*}[t]
\centering
\caption{Subject-level predictive validation using global duration and LZC features. Balanced accuracy and ROC-AUC are reported as mean $\pm$ standard deviation (SD) across repeated cross-validation folds. The best feature family within each dataset is bolded.}
\label{tab:subject-level-classification}
\scriptsize
\setlength{\tabcolsep}{3.8pt}
\renewcommand{\arraystretch}{1.12}
\resizebox{\textwidth}{!}{%
\begin{tabular}{llc*{3}{c}*{3}{c}}
\toprule
\multirow{2}{*}{Dataset} & \multirow{2}{*}{Task} & \multirow{2}{*}{$N$} & \multicolumn{3}{c}{Balanced accuracy} & \multicolumn{3}{c}{ROC-AUC} \\
\cmidrule(lr){4-6}\cmidrule(lr){7-9}
 & & & Hard & Trajectory & Combined & Hard & Trajectory & Combined \\
\midrule
ds004504 & AD vs HC & 65 & 0.806 $\pm$ 0.096 & 0.828 $\pm$ 0.081 & \textbf{0.839 $\pm$ 0.091} & 0.862 $\pm$ 0.089 & 0.854 $\pm$ 0.092 & \textbf{0.888 $\pm$ 0.083} \\
CAUEEG & AD vs HC & 500 & 0.704 $\pm$ 0.041 & 0.729 $\pm$ 0.039 & \textbf{0.739 $\pm$ 0.039} & 0.778 $\pm$ 0.042 & 0.790 $\pm$ 0.040 & \textbf{0.814 $\pm$ 0.037} \\
ds005385 & Young/old ($\leq$35 vs $\geq$60) & 323 & 0.689 $\pm$ 0.058 & 0.696 $\pm$ 0.057 & \textbf{0.703 $\pm$ 0.055} & 0.752 $\pm$ 0.059 & \textbf{0.777 $\pm$ 0.060} & 0.776 $\pm$ 0.059 \\
TD-BRAIN & Young/old ($\leq$35 vs $\geq$60) & 111 & 0.701 $\pm$ 0.095 & \textbf{0.752 $\pm$ 0.098} & 0.738 $\pm$ 0.098 & 0.800 $\pm$ 0.097 & \textbf{0.831 $\pm$ 0.095} & 0.827 $\pm$ 0.094 \\
\bottomrule
\end{tabular}%
}
\end{table*}

\section{Discussion}

This study asked whether temporal descriptors conventionally computed from hard label EEG microstate sequences can be made more sensitive to disease- and age-related variation by retaining the template evidence profile from which those labels are derived. Across four resting-state EEG datasets and multiple state model constructions, group-level \(K\)-means templates recovered canonical-like A--D structure at \(K=4\) and a finer-grained repertoire at \(K=7\). GEV was consistently high, indicating that the conventional fixed-template branch provided a stable reference model. Compared with this reference, the proposed evidence-aware descriptors showed stronger effects than their matched hard label counterparts. This pattern was most consistent for the duration family: high-evidence episode duration produced larger effects than conventional hard label duration across nearly all dataset--model combinations. Null state LZC also improved upon ordinary hard label LZC in most settings, although its benefit was more model-dependent and was attenuated for adaptive Leiden and Infomap models.

The signed effects observed in the present datasets indicated shorter global duration and higher symbolic complexity in AD and with increasing age. Thus, the main empirical finding is not that evidence-aware descriptors impose a particular physiological direction, but that they amplify the disease- and age-related differences already present in the matched hard label readout. In both AD--HC datasets and both adult-ageing datasets, the duration effects were consistently negative, whereas LZC and null state LZC effects were consistently positive. This cross-dataset consistency suggests that the proposed readouts captured a reproducible alteration of global topographic state dynamics within the present analysis pipeline.

The primary effect-size comparisons focused on global descriptors, including global mean duration and LZC, rather than state-specific measures tied to individual fixed-\(K\) templates. This choice was deliberate. Global descriptors summarise the overall temporal organisation of the topographic state sequence and are not tied to any particular template identity. They therefore provide a common readout across different \(K\) values and across both fixed state models and adaptive subject-specific Leiden or Infomap communities. State-specific duration, occurrence, and coverage remain useful for group-level fixed-\(K\) templates, where labels are aligned across participants, but they are less directly comparable for adaptive models in which state identities are subject-specific.

The strongest and most consistent result was the improvement of trajectory-derived duration descriptors. Conventional hard label duration measures the persistence of the winning label, but it does not indicate whether that persistence is supported by strong template evidence or by a sequence of weak argmax assignments. High-evidence episode duration addresses this limitation by measuring how each state's upper-tail evidence is organised in time. Because high-evidence samples were defined using within-state quantile thresholds, the marginal amount of high-evidence samples was controlled by construction. The resulting descriptor therefore does not estimate absolute occupancy. Instead, it quantifies whether high-confidence evidence occurs in sustained episodes or in fragmented bursts. In the present data, the shorter trajectory-duration effects suggest that AD and ageing were associated with more fragmented high-confidence evidence episodes, rather than simply with reduced time spent in any one template state.

These findings should not be interpreted as diminishing the value of conventional microstate parameters. The present analyses show that evidence-aware descriptors can increase sensitivity to disease- and age-related variation, but they do not replace the classical hard label parameter set. This is especially important for coverage, which remains a direct and interpretable measure of winner-state prevalence and has been reported as a disease- or cognition-related marker in previous AD and mild cognitive impairment (MCI) studies \citep{musaeus2019microstates,lian2021altered,lin2021differences,das2026electroencephalogram}. By construction, high-evidence episode metrics use within-state quantile thresholds and therefore do not estimate absolute state occupancy. They should therefore be interpreted as confidence-conditioned extensions of duration- and complexity-based readouts, while conventional coverage remains an essential marker of macro-scale state prevalence.

Null state LZC provided a second evidence-aware extension, targeting symbolic sequence complexity rather than state persistence. Ordinary LZC treats every hard label as a meaningful symbol, even when the corresponding topography is weakly explained by all available templates. By assigning samples with low template evidence to a dedicated null symbol, null state LZC preserves the conventional LZC computation while modifying the symbolic representation to encode periods of low confidence. In the present data, LZC effects were positive in AD and with increasing age, and null state LZC generally increased these effects relative to ordinary hard label LZC. This pattern is compatible with the duration result: if high-confidence state evidence is broken into shorter episodes and periods of low template evidence are explicitly represented, the resulting symbolic sequence can contain more diverse subsequences.

However, the benefit of null state LZC was not uniform across state model constructions. It was strongest for constrained fixed state models, where every GFP peak must be assigned to a limited template repertoire. In Leiden and Infomap models, the incremental benefit of null labelling was smaller. One explanation for this attenuation is that adaptive communities provide a larger, subject-specific topographic state repertoire. This reduces the number of samples that are poorly explained by all available states. The two approaches may therefore be complementary. When the state model is constrained, null state LZC provides an evidence-aware rejection mechanism. When the state model is individualised and sufficiently granular, the need for an explicit null symbol may be reduced.

The comparison between conventional group-level templates and adaptive topographic communities also clarifies the role of interpretability. At \(K=4\), group-level templates recovered canonical-like A--D patterns across datasets; at both \(K=4\) and \(K=7\), the aligned template sets supported state-specific measures such as duration, occurrence, and coverage. These descriptors are valuable when the scientific question requires cross-subject interpretation of specific aligned template states; canonical A--D labels are appropriate only for the \(K=4\) repertoire. In contrast, Leiden and Infomap produced subject-specific topographic communities with variable state numbers. These adaptive states were not interpreted as canonical microstates, but they provided individualised state repertoires from which label-invariant summary descriptors could be derived. The fact that evidence-aware readouts improved sensitivity across both conventional and adaptive models suggests that the main contribution arises from the readout of state evidence rather than from a particular clustering algorithm.

The predictive analyses also tested the practical utility of the descriptors. Across the AD--HC and age-related validation tasks, trajectory descriptors and combined feature sets containing both hard and trajectory descriptors often improved balanced accuracy or ROC--AUC relative to hard label features alone. These results suggest that evidence-aware descriptors contain information complementary to conventional microstate parameters. However, predictive performance should be interpreted conservatively unless all state model estimation, parameter selection, and feature selection steps are nested within cross-validation. The primary contribution of the present study is therefore not a final clinical classifier, but a measurement framework showing that retaining template evidence can increase the sensitivity of EEG topographic state descriptors.

The direction of the present duration and complexity effects should be interpreted against a heterogeneous literature. In AD, both shorter and longer microstate durations have been reported across studies and classes \citep{dierks1997eeg,strik1997decreased,das2026electroencephalogram}. Age-related findings also vary across states, age ranges, recording conditions, and frequency bands, with both increases and decreases reported across the lifespan \citep{tomescu2018from,zanesco2020within,jabes2021resting,huang2024magnetoencephalography}. Findings for sequence complexity are similarly mixed: reduced microstate LZC has been reported in AD \citep{tait2020eeg}, whereas another study found higher microstate sequence LZC in AD than in healthy controls \citep{wan2024beyond}. The shorter duration and higher LZC observed here therefore fall within previously reported directions, but neither appears to be a universal signature of AD or ageing. The present framework increases sensitivity to phenotype-related variation without resolving this directional heterogeneity.

Several limitations should therefore be emphasised. The most important is that the direction of EEG microstate duration and complexity effects remains method-dependent. The present results indicate shorter duration and higher LZC across four datasets under an alpha-band pipeline based on GFP peaks and template evidence trajectories. This should not be interpreted as a universal direction of AD- or ageing-related microstate change. Preprocessing choices, including artefact rejection, band selection, GFP peak extraction versus full-time-series fitting, template normalisation, and polarity handling, may influence the resulting sequence and the sign of duration or complexity effects. The same applies to backfitting strategy, temporal smoothing, clustering algorithm, the selected number of states, group-level versus subject-specific templates, and null state or threshold definitions. Therefore, the evidence-aware readout should be viewed as improving measurement sensitivity within a specified pipeline, not as providing a pipeline-invariant physiological direction. Second, the analyses were conducted in sensor space and focused on alpha-band GFP peak topographies. The proposed framework should be tested in other frequency bands, broadband or full time-series settings, and source-space EEG or MEG representations. Third, high-evidence episode metrics depend on quantile thresholds and evidence-sharpening choices. Although threshold families and parameter sweeps reduce the risk of single-threshold overfitting, future work should evaluate the reliability and physiological interpretation of these parameters more systematically. 

Future work should directly benchmark the sources of directional variability. A useful next step would be to apply multiple preprocessing pipelines, frequency bands, segmentation methods, template-fitting strategies, smoothing settings, and state-number choices to the same datasets, while reporting both signed and absolute effects for hard label and evidence-aware descriptors. Such analyses would help determine whether shorter duration and higher LZC reflect a robust feature of particular AD or ageing cohorts, a consequence of alpha-band GFP peak dynamics, or an interaction between disease effects and state model construction. Standardised reporting of signed effects, template-fitting procedures, and sequence-construction choices will be essential for comparing results across studies.

A broader future direction is to examine whether the evidence-preservation principle examined here can inform emerging EEG representation-learning frameworks. Recent EEG foundation models aim to learn transferable representations across heterogeneous datasets and clinical settings, while microstate-informed approaches have begun to use microstate sequences as structured representations of EEG dynamics \citep{jiang2024large,wang2024eegpt,wang2025lead,nguyen2025transforming,tian2025atoms}. The present results suggest that retaining graded template evidence, rather than relying exclusively on discrete microstate labels, may provide additional information for such approaches. Future studies could therefore investigate whether evidence-aware microstate representations improve generalisation or interpretability in larger-scale EEG models.

In summary, conventional hard label microstate metrics remain valuable summaries of state prevalence and persistence of the winning label. However, they discard information about template confidence, state competition, and periods of low template evidence. By retaining the full template evidence trajectory, the proposed framework provides confidence-aware extensions of duration and LZC that more sensitively capture disease- and age-related variation in EEG topographic state dynamics. The present data show shorter global duration and higher symbolic complexity in AD and with increasing age, but these signed effects should be interpreted as pipeline-specific and hypothesis-generating until tested under standardised, multimodal, and longitudinal designs.

\section{Conclusion}

This study examined whether temporal descriptors derived from EEG topographic states can be made more sensitive by retaining the template evidence profile that is discarded during winner-take-all assignment. Across four resting-state EEG datasets and multiple fixed and adaptive state model constructions, evidence-aware duration descriptors consistently produced stronger disease- and age-related effects than matched hard label duration, while null state LZC improved sensitivity in most settings. These findings do not replace conventional microstate parameters, particularly coverage, but they show that hard labels need not be the only readout of an underlying state model. The observed shorter duration and higher symbolic complexity in AD and with increasing age should be interpreted within the specific analysis pipeline rather than as universal physiological directions. Overall, retaining graded state evidence provides a practical extension of conventional microstate analysis for characterising EEG topographic dynamics.

\section{Data and Code Availability}\label{sec:data-code-availability} \textbf{Data availability.} The data analysed in this study were obtained from third-party repositories and remain subject to the access conditions of their original providers. OpenNeuro ds004504 (snapshot 1.0.9) is available at \url{https://openneuro.org/datasets/ds004504/versions/1.0.9}, and OpenNeuro ds005385 is available at \url{https://openneuro.org/datasets/ds005385}. CAUEEG is available through its official repository and access procedure at \url{https://github.com/ipis-mjkim/caueeg-dataset}. TD-BRAIN V3.1 is available from the Brainclinics Foundation subject to its data-use agreement at \url{https://brainclinics.com/resources/tdbrain-dataset/introduction}. Raw EEG data are not redistributed with this article or the code repository. \textbf{Code availability.} Analysis code and documentation are openly available at \url{https://github.com/OnismKD/evidence-trajectory-microstates} (version 0.1.0) under the BSD 3-Clause License. The repository includes the analysis implementation, dataset-specific configuration files, a locked computational environment, tests, and a workflow that verifies the ds004504 classification results reported in Table~\ref{tab:subject-level-classification}. Dataset files and restricted participant metadata are not included; researchers must obtain them from the original providers and comply with the corresponding access and use terms. 

\section{Acknowledgement}\label{sec:Acknowledgement}
Kaidong Wu and Fei He were supported by the Engineering and Physical Sciences Research Council (EPSRC) grant [EP/X020193/1]. We gratefully acknowledge the investigators, clinical/research staff, and participants who contributed to the TD-BRAIN database, OpenNeuro ds004504, and OpenNeuro ds005385. We also acknowledge OpenNeuro for hosting the ds004504 and ds005385 datasets, and the Brainclinics Foundation/Synapse infrastructure for providing access to TD-BRAIN. The original data contributors and repositories bear no responsibility for the analyses, interpretations, or conclusions presented here.


\section{Ethics Statement}

This study used only previously collected, de-identified datasets obtained from established public or controlled-access repositories. No new participants were recruited and no new human data were collected; therefore, additional ethical approval was not required for the present secondary analyses. Ethical approval and informed consent for the original data collection were obtained by the respective dataset providers, as described in the original dataset publications.

\section{Conflict of Interest Statement}

The authors declare no conflicts of interest.

\bibliographystyle{plainnat}
\bibliography{references}

\end{document}